# Enhancing Optical Performance of Liquid Crystal Lens Arrays via Electrode Design Optimization

RUI DING, LI-LAN TIAN *, YI ZHENG, YU-MENG ZENG, FAN ZOU, YUE NIU, RONG-FU LIU, AND JI-WEI ZHOU

*College of Physics, Chengdu University of Technology, Chengdu 610059, China*
*tianlilan@cdut.edu.cn*

**Abstract:** A liquid crystal (LC) lens array based on double-layer composite electrodes, characterized by a large aperture, short focal length, and low operating voltage is demonstrated. The lens array consists of an LC layer, a top common electrode, a bottom double-layer composite electrode layer, and an oxide layer. The bottom double-layer composite electrode layer comprises the pixel electrodes and the auxiliary electrode. In focusing mode, the pixel electrodes receive operational voltage to establish the LC layer's electric field, with the auxiliary electrode applying reduced voltage for field optimization. Experiment results show that the proposed LC lens array achieves the shortest focal length of 3.3 mm when the pixel electrodes are set at 5.2 $V_{rms}$ and the auxiliary electrode is set at 2.6 $V_{rms}$. This design addresses the technical challenge of achieving larger apertures (800 μm or more), offering enhanced viewing zones and improved 3D performance. This configuration provides an ideal refractive index distribution in a relatively thick LC layer, enabling 2D/3D switchable display with performance superior to current LC lens arrays of equivalent aperture. Furthermore, the proposed structure demonstrates excellent tolerance to manufacturing errors.

## 1. Introduction

In recent years, augmented reality (AR) and virtual reality (VR) technologies have developed rapidly, with applications expanding into education [1, 2], healthcare [3, 4], industrial design [5, 6], and entertainment [7]. Pioneering products like Apple's Vision Pro, which integrates high-resolution micro-display technology with 3D depth perception, are continuously pushing the limits of immersive experiences [8]. Research indicates that the immersion of AR/VR systems is closely related to 3D depth perception, which simulates physiological mechanisms such as binocular disparity and accommodation convergence in the human visual system, significantly enhancing users' spatial perception and interaction realism in virtual environments [9, 10]. Consequently, the development of display technologies that can achieve high-precision 3D depth perception has become a critical challenge in advancing the performance of AR/VR systems.

Currently, 3D display technologies are mainly divided into two categories: traditional stereoscopic displays that rely on glasses and autostereoscopic displays that require no external devices [11]. Traditional stereoscopic displays achieve 3D effects by separating left and right eye images through time-division or polarization techniques. However, time-division methods, such as shutter-based 3D displays, can lead to flickering due to rapid image switching, while polarization techniques, such as polarized 3D displays, may suffer from crosstalk caused by imperfect polarization filtering [12-14]. In contrast, autostereoscopic display systems directly guide light to achieve 3D effects without glasses, offering significant advantages in viewing comfort, resolution, and user freedom [15]. Traditional autostereoscopic display technologies primarily rely on fixed parallax barriers [16, 17] and lenticular lenses [18, 19], both of which exhibit performance limitations when displaying 2D content [20, 21]. This makes seamless 2D/3D switching a key research direction. In comparison, LC lens arrays, with their electrically tunable refractive index and lack of mechanical structures, have emerged as a novel solution for 2D/3D switchable display technology.

Several 2D/3D switchable display technologies based on LC lens arrays have been proposed. For example, switchable barrier technology, based on the voltage response characteristics of crossed polarizers and LC layers, achieves high-brightness full-area transmission in 2D mode with a simple structure and strong compatibility. However, in 3D mode, the selective blocking of pixel light by the parallax barrier results in significant brightness attenuation and limited viewing angles, severely impacting the immersive experience [22-24]. Curved electrode lenses quickly switch between non-focusing and focusing modes by matching the LC refractive index with the concave structure through voltage adjustment, offering fast response times. However, their manufacturing process is complex, operating voltages are high, and even with increased voltage, it is difficult to adjust the focal length to infinity [25-27]. Polarization-switching lenses innovatively combine fixed anisotropic lenses with polarization modulation layers, using polarization state conversion to control refractive index differences, offering high optical efficiency and low energy consumption. However, their dependence on the polarization state of incident light can lead to instability in 3D effects under non-polarized light sources or ambient light interference [28-30]. Furthermore, current liquid crystal lens arrays typically feature unit pitches ranging from 100 to 400 micrometers, which limits the number of pixels each lens can cover, making it difficult to achieve more viewing zones and higher-quality 3D displays. Developing liquid crystal lens arrays with larger apertures, such as 800 micrometers or more, remains a significant technical challenge in the field [31-34]. In summary, although existing technologies have optimized 2D/3D switching in various dimensions, addressing the trade-off between brightness and resolution, improving environmental adaptability and manufacturing yield, and achieving larger apertures remain critical bottlenecks for commercialization.

This paper demonstrates a large-aperture LC lens array based on double-layer composite electrodes for 2D/3D switchable display. The lens array comprises an LC layer, a top common electrode, a bottom double-layer composite electrode layer and an oxide layer. The bottom double-layer composite electrode layer comprises the main pixel electrodes (referred to as the pixel electrodes) and an auxiliary pixel electrode (referred to as the auxiliary electrode). The combination of pixel and auxiliary electrodes optimizes the electric field distribution, creating a more pronounced gradient electric field in the LC layer. This configuration enhances the parabolic distribution of the refractive index in the LC layer, resulting in a larger refractive index contrast and consequently a shorter focal length. Experiment results show, with operating voltages of 5.2 $V_{rms}$ for the pixel electrodes and 2.6 $V_{rms}$ for the auxiliary electrode, the focal length of the LC lens array can be reduced to 3.3 mm while maintaining a large aperture of 800 μm. Compared to existing technologies, the proposed LC lens array not only simplifies the structure but also significantly reduces manufacturing complexity, offering a high-performance, low-cost solution for 2D/3D switchable display.

## 2. Device structure and principle

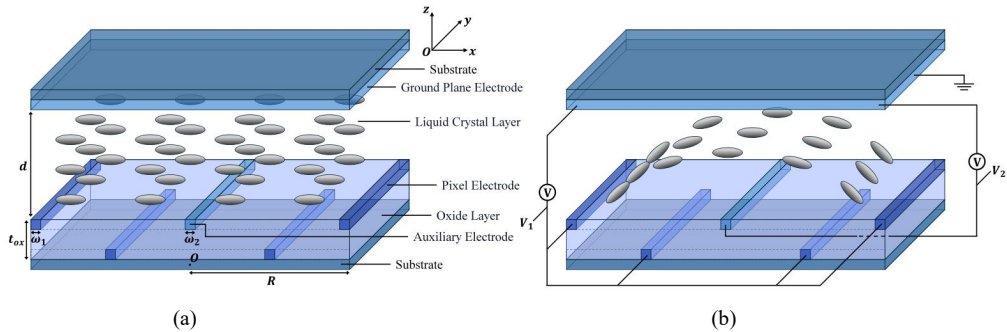

**Fig. 1.** Structure of a single LC lens array in the (a) non-focusing mode and (b) focusing mode.

Fig. 1 shows the side view of the LC lens array in the (a) non-focusing mode and (b) focusing mode. To fabricate the LC lens array, a planar electrode is deposited on the lower surface of the top glass substrate as the ground electrode, while five strip electrodes with the same configuration are deposited on the upper and lower surfaces of the oxide layer at the bottom of the lens structure in an alternating pattern. The central strip electrode on the upper surface of the oxide layer serves as the auxiliary electrode, while the four strip electrodes at the ends of the upper surface and the 1/4 and 3/4 positions of the lower surface serve as the pixel electrodes. Polyimide films are coated on the lower surface of the top glass substrate and the upper surface of the oxide layer to initialize the alignment of the LC in the LC layer.

As shown in Fig. 1(a), when no voltage is applied to the electrodes, the LC lens array operates in the non-focusing mode, with the LC molecules aligned horizontally along the x-axis. When light is incident perpendicular to the glass substrate, the LC layer exerts the same refractive effect on the light, and the lens array does not focus the light. As depicted in Fig. 1(b), when voltage is applied to both the pixel and auxiliary electrodes, the LC lens array enters the focusing mode, with the LC molecules aligned at varying angles due to the non-uniform electric field. When light is incident perpendicularly, the LC molecules in different regions of the LC layer exert different refractive effects on the light, resulting in a focusing effect.

As shown in Fig. 1(a), $R$ represents the radius of a single LC lens array, $d$ represents the thickness of the LC layer, and $\omega_1$, $\omega_2$ correspond to the widths of the pixel and auxiliary electrodes, respectively. All electrodes have the same thickness. As shown in Fig. 1(b), in the focusing mode, the top common electrode is grounded, while the bottom pixel and auxiliary electrodes are applied with voltages $V_1$ and $V_2$, respectively.

Since the LC material used is positive, the effective refractive index of linearly polarized light incident along the x-axis can be expressed as [35,36]:

$$n_{eff}(\theta) = \frac{n_o n_e}{\left(n_o^2 sin^2\theta + n_e^2 cos^2\theta\right)^{1/2}} \quad (1)$$

where $\theta$ represents the tilt angle between the LC molecules and the x-axis, $n_o$ represents the ordinary refractive index when the light wave vibration direction is perpendicular to the long axis of the LC molecules, and $n_e$ represents the extraordinary refractive index when the light wave vibration direction is parallel to the long axis of the LC molecules.

Using equation (1), the cumulative average effective refractive index distribution along the x-axis of the LC lens array can be calculated. The phase delay distribution of the light beam relative to the incident plane at the corresponding position can be expressed as [35]:

$$\phi = \frac{2\pi}{\lambda} \int_0^d n_{eff}(\theta) dz \quad (2)$$

where $d$ represents the thickness of the LC layer.

According to the Fresnel approximation, the focal length of the LC lens array can be expressed as [37, 38]:

$$f = \frac{R^2}{2d\delta n} \quad (3)$$

where $\delta n$ represents the difference in effective refractive index between the center and edge of the LC lens array.

From equation (3), it can be seen that to achieve a shorter focal length, the value of $\delta n$ should be increased. The most direct way is to increase the voltage on the pixel electrodes to make the LC molecules at the edge of the LC layer rotate more significantly, thereby reducing the edge refractive index. However, the voltage on the auxiliary electrode must be adjusted to maintain the refractive index distribution curve of the LC layer close to the ideal parabolic curve.

## 3. Simulation results and discussion

To optimize the parameters used in the LC model, a series of numerical calculations were performed using Python and Tech-Wiz LCD 3D software from Sanayi System Co., Ltd. The LC material parameters used in the simulation are listed in Table 1. In the simulation, light with a wavelength of 550 nm was used. The optimized and calculated parameters of the LC lens array are listed in Table 2. In this configuration, the height of all electrodes is uniformly set to 0.04 μm.

The calculation principle is based on the extended Jones matrix. The Jones matrix of polarized light propagating along the z-axis in space is expressed as [39,40]:

$$\vec{E} = \begin{pmatrix} E_x \\ E_y \end{pmatrix} = \begin{pmatrix} E_{0x}e^{i\phi_x} \\ E_{0y}e^{i\phi_y} \end{pmatrix} \tag{4}$$

where $E_{0x}$ and $E_{0y}$ represent the amplitudes of polarized light propagating along the x-axis and y-axis, respectively, and $\phi_x$ and $\phi_y$ represent the phases of polarized light in the two coordinate directions.

Table 1. Material parameters of LC lens array.

| LC material | $\varepsilon_\parallel$ | $\varepsilon_\perp$ | $K_{11}$ | $K_{22}$ | $K_{33}$ | $n_e$ | $n_o$ |
|---|---|---|---|---|---|---|---|
| E7-LC | 16.7 | 5.3 | 11.7 pN | 8.8 pN | 11.5 pN | 1.741 | 1.517 |

Table 2. Parameters of ICE-LC lens array.

| Parameters of the LC lens array | Value |
|---|---|
| Lens pitch ($2R$) | 800 μm |
| Pixel electrode width ($\omega_1$) | 20 μm |
| Auxiliary electrode width ($\omega_2$) | 20 μm |
| Cell gap ($d$) | 150 μm |
| Oxide layer thickness ($t_{ox}$) | 80 μm |
| Maximum refractive index ($n_{max}$) | 1.741 |
| Minimum refractive index ($n_{min}$) | 1.578 |
| Focal length ($f$) | 3.27 mm |

Fig. 2 illustrates the electric potential and electric field distribution of the proposed LC lens array at $V_1 = 5$ $V_{rms}$ and $V_2 = 2.5$ $V_{rms}$ with the electric field and potential distribution of the LC layer in a single LC lens array shown in the red dashed box. From Fig. 2(a), the potential distribution becomes progressively denser from the center to the edge of the lens, indicating a corresponding increase in electric field strength. This gradually increasing electric field strength helps to make the rotation angle of the LC molecules from the center to the edge smoother, resulting in a smooth refractive index distribution. Fig. 2(b) further demonstrates that the electric field strength increases from the center to the edge of the lens, consistent with the potential distribution observed in Fig. 2(a). This form of potential and electric field distribution is the basic condition for the LC lens to achieve focusing effect.

To further study the light refraction characteristics of the LC lens array, this paper analyzed the orientation distribution of the LC molecules. Therefore, this paper simulated the orientation angle of the LC molecules in the LC lens array under the conditions of $V_1 = 5$ $V_{rms}$ and $V_2 = 2.5$ $V_{rms}$, and showed the cross-sectional view along the x-axis and the top view along the z-axis, as shown in Fig. 3.

From Fig. 3(a), it can be seen that the LC molecules at the center of the LC layer remain horizontally aligned, while in the regions near the edge, the rotation angle of the LC molecules gradually increases, causing their alignment angle relative to the horizontal plane to rise. At the outermost edge of the lens unit, the LC molecules reach their maximum alignment angle. According to equation (1), as the angle increases, the refractive index of light gradually decreases. Therefore, the LC molecules at the center of the LC layer always exhibit the maximum refractive index (1.741), while those at the edge correspond to the minimum refractive index (1.578 under the simulation conditions in Fig. 3). This difference in refractive index between the center and edge of the lens is the basic condition for the LC lens to achieve focusing effect.

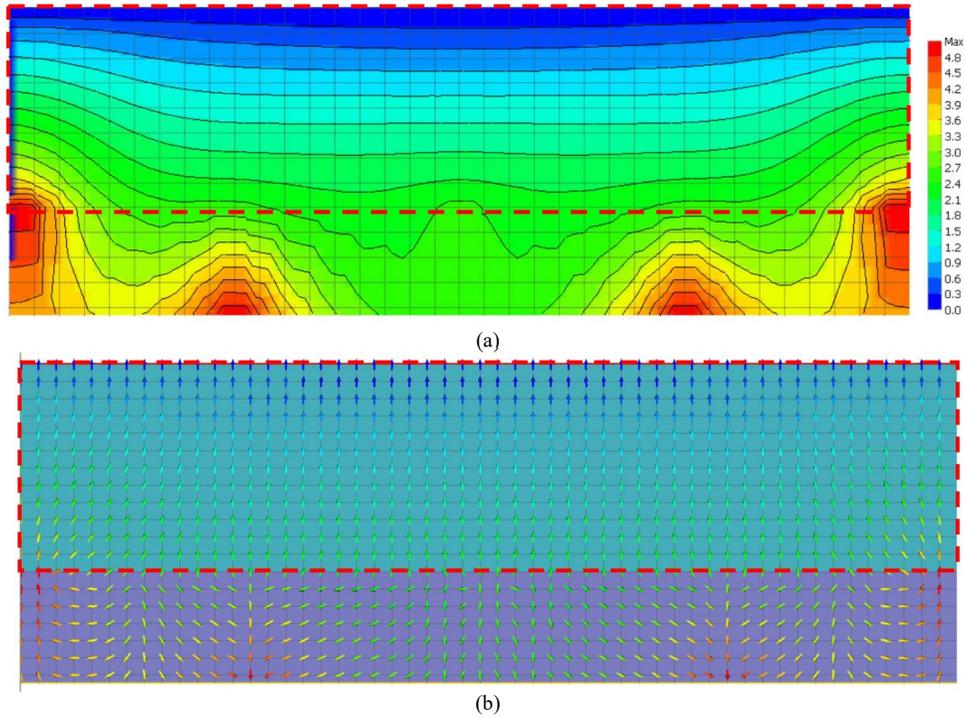

**Fig. 2.** Simulated (a) electric potential and (b) electric field distribution of the proposed LC lens array at $V_1 = 5$ V$_{rms}$ and $V_2 = 2.5$ V$_{rms}$.

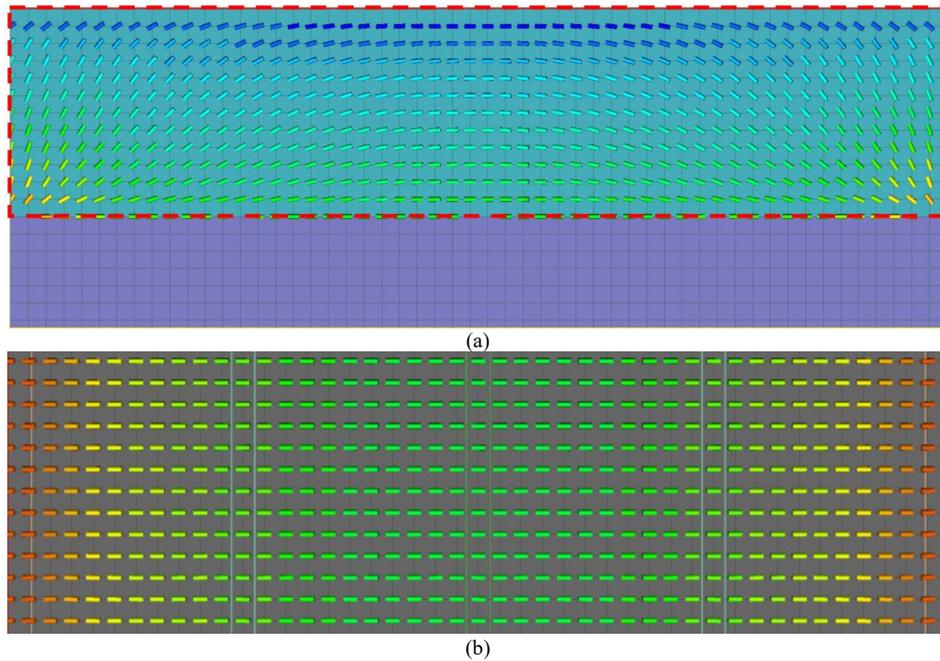

**Fig. 3.** Simulated (a) top view and (b) cross-section of the LC molecules orientation in the LC layer of the LC lens

array at $V_1 = 5$ $V_{rms}$ and $V_2 = 2.5$ $V_{rms}$.

In addition, this paper simulated the top view of the orientation distribution of the LC molecules, as shown in Fig. 3(b). From the top view, it can be observed that the angle of the LC molecules at the center of the lens is almost parallel to the horizontal plane, while the orientation angle of the LC molecules on both sides increases with the distance from the center, showing a trend consistent with Fig. 3(a). It is worth noting that according to equation (1), if the LC molecules at the edge are perpendicular to the horizontal plane, the edge refractive index of the lens unit will reach the minimum value, and according to equation (3), the focal length will also reach the minimum value, helping to achieve the shortest focal length. However, due to the anchoring energy between the inner surfaces of the upper and lower substrates of the LC lens unit, the LC molecules at the edge of the LC lens unit are almost impossible to reach a state completely perpendicular to the horizontal plane. Therefore, to achieve a shorter focal length while maintaining good focusing performance, the operating voltage cannot simply be increased.

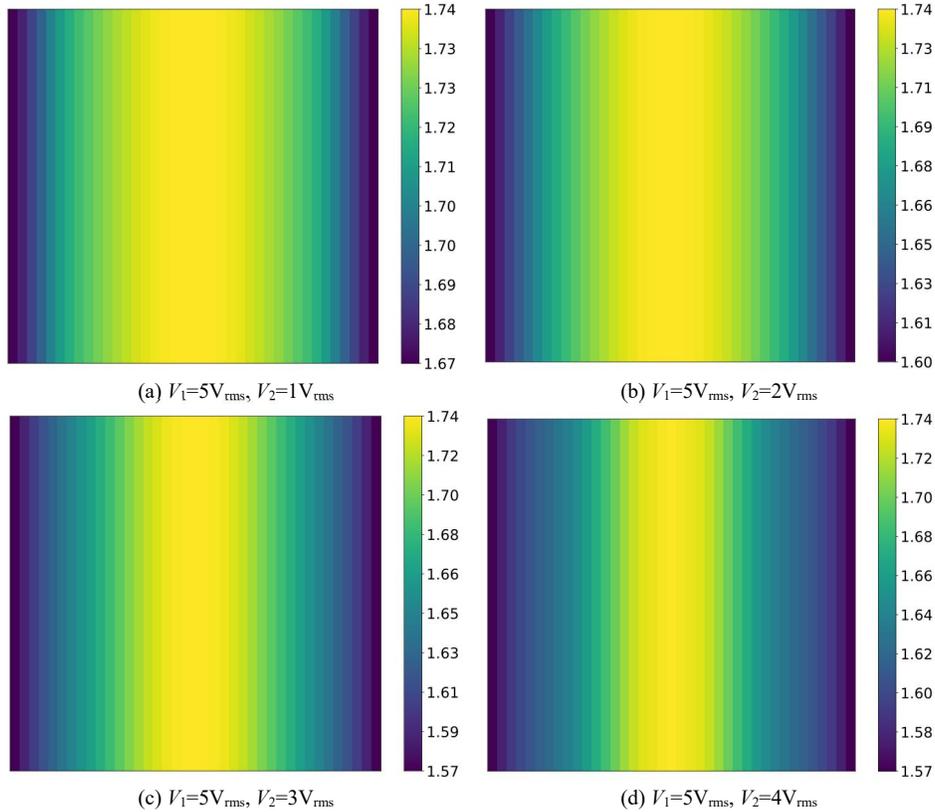

**Fig. 4.** Simulated (a)-(d) top view of the equivalent refractive index distribution along the *x*-axis with different with different $V_1$ and $V_2$ voltages.

Fig. 4(a)-(d) shows the top view of the equivalent refractive index spatial distribution of the LC lens array unit under fixed $V_1 = 5$ $V_{rms}$ and different $V_2$ voltages. Fig. 5(a)-(d) presents the effective refractive index distribution curve along the x-axis of the LC lens array unit under fixed $V_1 = 5$ $V_{rms}$ and different $V_2$ voltages. From Fig. 4, it is evident that the spatial distribution of the refractive index is symmetric about the centerline of the lens. When $V_2$ is 1 $V_{rms}$, 2 $V_{rms}$, and 3 $V_{rms}$, the refractive index distribution of the LC lens unit shows a relatively good gradient change. However, when $V_2 = 4$ $V_{rms}$, the refractive index distribution of the LC lens unit shifts significantly, failing to maintain the original gradient change. From Fig. 5, it can be more clearly see the change in refractive index distribution with $V_2$ voltage. In Fig. 5(c)

and Fig. 5(d), although higher $V_2$ voltages result in a larger effective refractive index difference, the rotation amplitude of the LC molecules at the edge of the lens is significantly smaller than that of the molecules near the center due to the limitation imposed by anchoring energy. As a result, the refractive index distribution shows a sharply rising trend, deviating from the standard parabolic distribution and failing to achieve ideal focusing effect. This is the specific manifestation of the earlier statement that "the operating voltage cannot simply be increased". As shown in Fig. 6, when $V_1 = 5$ $V_{rms}$ and $V_2 = 2.5$ $V_{rms}$, the maximum refractive index difference of 0.1631 is obtained, ensuring the refractive index distribution optimally matches the standard parabolic distribution.

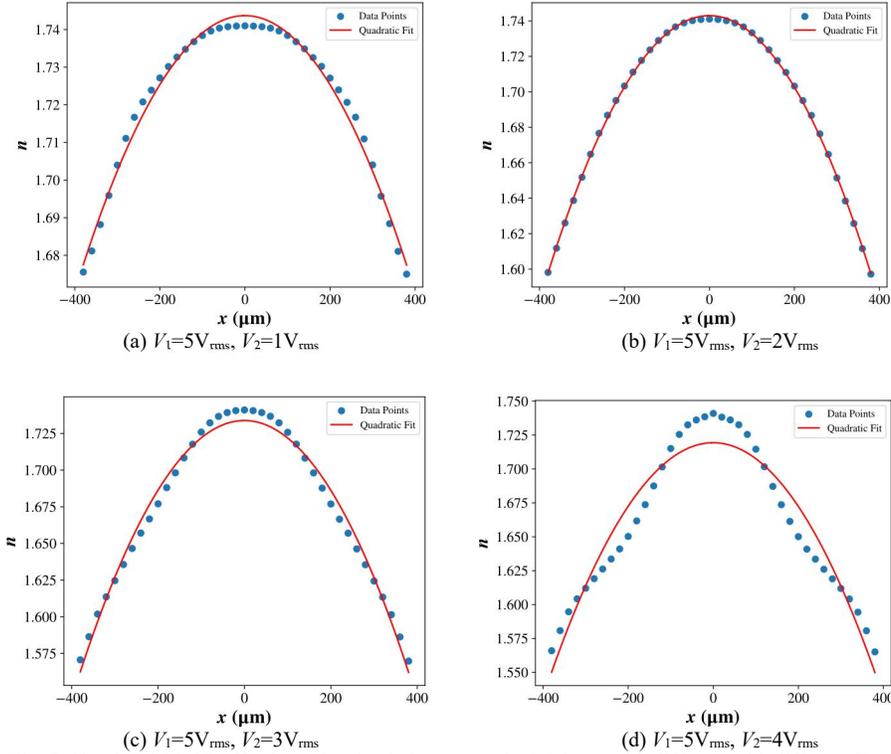

**Fig. 5.** Simulated (a)-(d) effective refractive index across the LC layer with different with different $V_1$ and $V_2$ voltages.

In fact, according to the simulation, the refractive index distribution of the proposed LC lens array can match the parabolic distribution well when $V_1 = 5$ $V_{rms}$ and $V_2 \leq 2.5$ $V_{rms}$. The fact that the refractive index distribution can match the standard parabolic distribution well largely indicates that this structure of the LC lens array can achieve the best focusing effect under these operating voltages.

As shown in Fig. 7, this paper fixed the voltage of the pixel electrodes at 5 $V_{rms}$ and varied the voltage $V_2$ on the auxiliary electrode to obtain refractive index distribution curves at different voltages. From the figure, it can be observed that as the $V_2$ gradually decreases, the upper edge of the refractive index distribution becomes flatter, and the refractive index difference between the center and edge of the LC lens unit progressively decreases. This indicates that as the voltage decreases, the LC lens gradually loses its focusing ability. Therefore, the primary role of the auxiliary electrode is to assist the pixel electrodes in adjusting the orientation of the LC molecules near the center of the LC layer, ensuring the refractive index distribution of the LC lens unit better approximates a parabolic shape, thereby maintaining optical focusing performance.

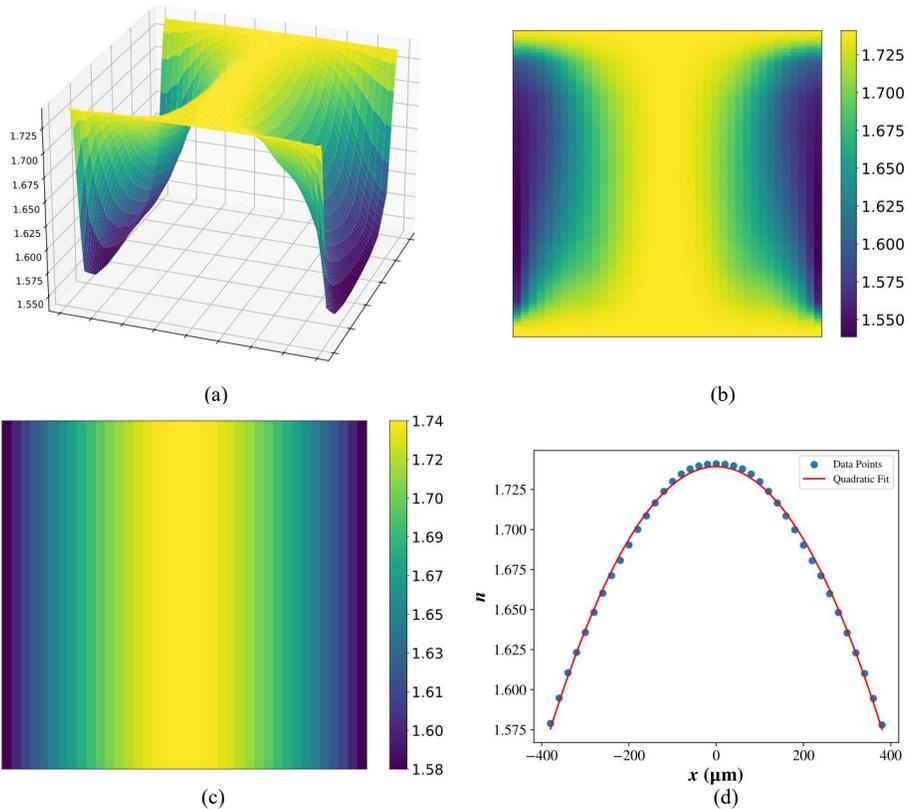

**Fig. 6.** LC lens array unit at $V_1$ = 5 $V_{rms}$ and $V_2$ = 2.5 $V_{rms}$: (a) 3D refractive index distribution, (b) refractive index top view, (c) equivalent refractive index top view, (d) refractive index distribution curve.

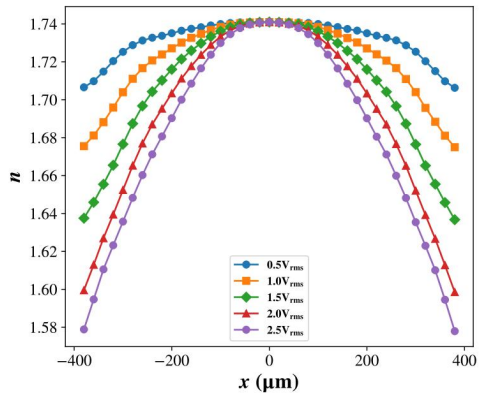

**Fig. 7.** Simulated effective refractive index curve across an individual LC lens array at different $V_1$.

   To further study the role of the auxiliary electrode in the lens structure, this paper removed the auxiliary electrode at the center of the LC lens unit and obtained the simulation model depicted in Fig. 8(a), fitting its refractive index distribution. As shown in Fig. 8(b), the observed refractive index distribution curve does not reach the ideal parabolic shape. This deviation is due to the limited penetration effect of the pixel electrodes at the edge of the lens, which cannot orient the LC molecules at the center of the lens. This results in a refractive index distribution curve that is too flat at the top, preventing focusing. The introduction of the auxiliary electrode compensates for this deficiency, further emphasizing the indispensable role of the auxiliary electrode in achieving effective optical focusing in the proposed LC lens structure.

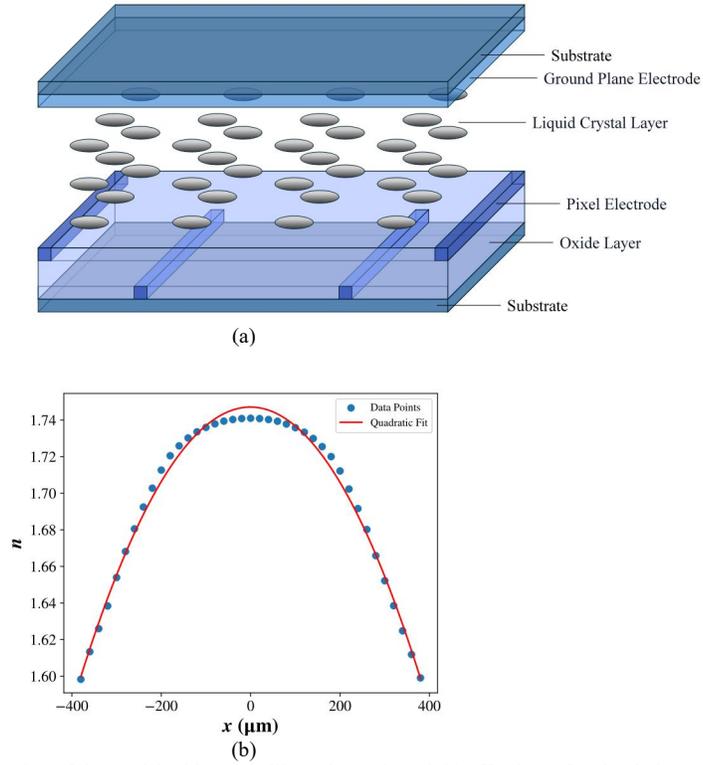

**Fig. 8.** Simulated (a) configuration of the model without auxiliary electrode and (b) effective refractive index curve across the LC lens array at $V_1 = 5$ $V_{rms}$ of the model without auxiliary electrode.

In industrial mass production, issues such as manufacturing errors are often considered. Therefore, the potential errors that may occur in the proposed LC lens array during industrial production need to be analyzed. Here, this paper only discusses a single LC lens array unit. Due to manufacturing processes, embedding strip electrodes on the upper and lower surfaces of the oxide layer may result in a misalignment in the $x$-axis direction, typically ranging from 0 to 5 $\mu m$. Therefore, this paper simulates four different cases of electrode misalignment, as shown in Fig. 9. From Fig. 9(a), it can be seen that the four parabolic curves maintain a good parabolic distribution on a macroscopic level, indicating that within the error range introduced by manufacturing processes, the LC lens can still achieve good focusing performance. However, as shown in Fig. 9(b), on a more microscopic scale, the tops of the refractive index curves of the four groups are slightly offset compared to the reference group with the standard parabolic curve. Nevertheless, such small offsets are insufficient to cause macroscopic phase changes, meaning that the manufacturing errors have minimal impact on the imaging performance of the proposed lens structure.

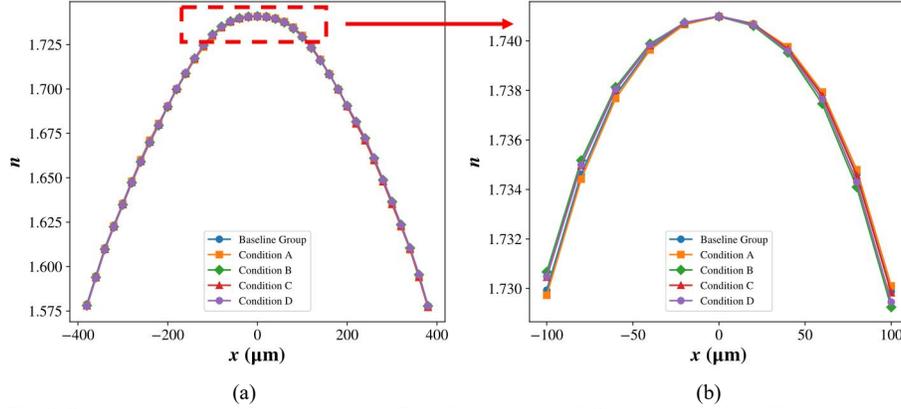

(a)                       (b)

**Fig. 9.** Simulated relative refractive index profiles of an individual LC lens array under different electrode misalignment conditions at $V_1 = 4.5$ $V_{rms}$ and $V_2 = 0.9$ $V_{rms}$. Condition A: Auxiliary electrode shifted left by 5 μm, Condition B: Auxiliary electrode and two-pixel electrodes on the lower surface of the oxide layer shifted left by 5 μm, Condition C: Left pixel electrode on the lower surface of the oxide layer shifted left by 5 μm, Condition D: Two-pixel electrodes on the lower surface of the oxide layer shifted left by 5 μm.

## 4. Experiment results and discussion

Fig. 10 shows the fabricated LC lens array with a size of 20 mm ×20 mm and the pitch of the LC array is 800μm. To investigate the interference characteristics of the LC lens array, parallel green light is employed as the incident beam. To enhance the visibility and contrast of the interference fringes, a linear polarizer is attached to the output side of the LC lens array, with its transmission axis aligned parallel to the initial orientation direction of the LC molecules. Figure 10 presents the interference fringes observed at (a) $V_1 = 0$ $V_{rms}$, $V_2 = 0$ $V_{rms}$, (b) $V_1 = 3.4$ $V_{rms}$, $V_2 = 1.7$ $V_{rms}$, and (c) $V_1 = 5.2$ $V_{rms}$, $V_2 = 2.6$ $V_{rms}$, respectively, with the recording plane positioned 1 mm behind the lens array. As shown in Fig. 10, the number of interference fringes is increasing with increasing voltage, indicating that the phase modulation capability of the LC lens array is strengthened. At $V_1 = 5.2$ $V_{rms}$ and $V_2 = 2.6$ $V_{rms}$, the LC lens array produces the most distinct and well-defined interference pattern.

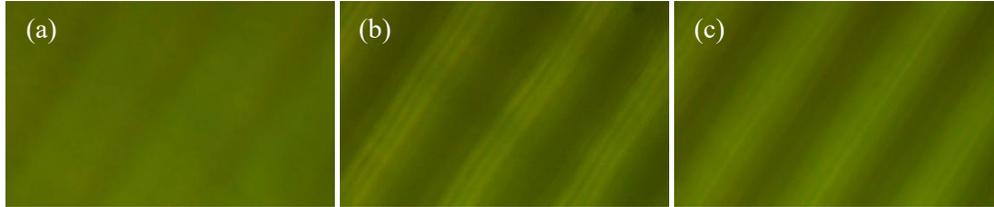

**Fig. 10.** The interference fringes of the LC array at (a) $V_1 = 0$ $V_{rms}$ and $V_2 = 0$ $V_{rms}$, (b) $V_1 = 3.4$ $V_{rms}$ and $V_2 = 1.7$ $V_{rms}$, (c) $V_1 = 5.2$ $V_{rms}$ and $V_2 = 2.6$ $V_{rms}$, respectively.

The LC lens array demonstrates voltage-tunable focusing characteristics when illuminated by parallel green light. By integrating a linear polarizer with its polarization axis aligned to the LC director orientation, the focal line contrast is significantly enhanced. Experimental results reveal that increasing the applied voltage progressively refines and brightens the focal line, indicating a nonlinear reduction in effective focal length as shown in Fig.11. At $V_1 = 5.2$ $V_{rms}$, $V_2 = 2.6$ $V_{rms}$, the shortest focal length (3.3 mm) is achieved, matching the theoretical value (3.27mm). The LC lens array exhibits advantages including low driving voltage, simple fabrication, making it promising for dynamic optical applications.

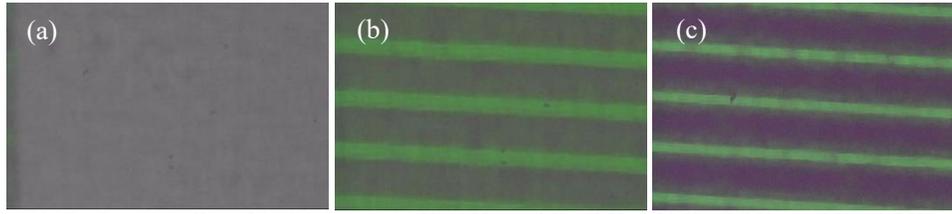

**Fig. 10.** The Focusing effect of the LC array at (a) $V_1 = 0$ V$_{rms}$ and $V_2 = 0$ V$_{rms}$, (b) $V_1 = 3.4$ V$_{rms}$ and $V_2 = 1.7$ V$_{rms}$, (d) $V_1 = 5.2$ V$_{rms}$ and $V_2 = 2.6$ V$_{rms}$, respectively.

Fig. 11 shows the focal length of the LC lens array measurement platform, and the relationship curve between the lens focal length and the operating voltage of the auxiliary electrode, where the horizontal axis represents the voltage $V_2$ applied to the auxiliary electrode, and $V_1$ is always 5.2 V$_{rms}$. When no voltage is applied, the focal length tends to infinity, and the LC lens has no focusing effect. As the applied voltage increases, the focal length decreases sharply due to the inverse relationship between focal length and refractive index difference in equation (3). When the applied voltage reaches $V_2 = 1.5$ V$_{rms}$, the focal length is 5.1mm. However, when the voltage exceeds this value, the degree of focal length reduction decreases. This is because as the applied voltage increases, the interaction between the potential energy and anchoring energy on the inner surfaces of the upper and lower substrates of the LC lens unit gradually saturates, making it difficult for the LC molecules to rotate further with increasing voltage. When $V_1 = 5.2$ V$_{rms}$ and $V_2 = 2.6$ V$_{rms}$, the focal length of the LC lens array reaches its minimum value of 3.3 mm.

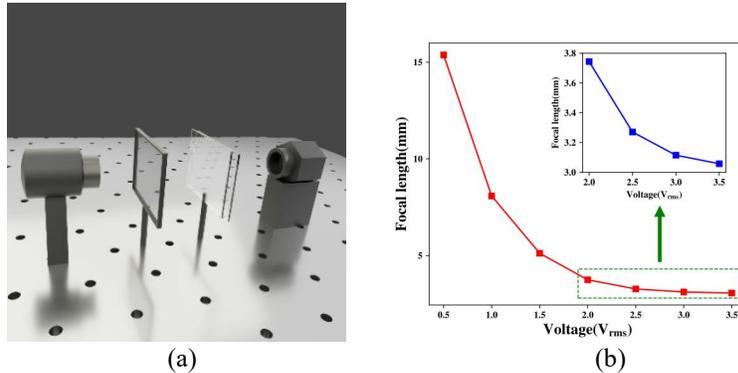

(a)          (b)

**Fig. 11.** (a) The focal length of the LC lens array measurement platform, and (b) the voltage-dependent focal length of an individual LC lens.

## 5. Conclusion

An LC lens array with double-layer composite electrodes for 2D/3D switchable displays is demonstrated. The LC lens array features a simple structure, large aperture, short focal length, and low operating voltage. By combining main and auxiliary pixel electrodes, the design achieves a parabolic refractive index distribution, reducing the focal length and driving voltage. Experiment results show that with the pixel electrodes at 5.2 V$_{rms}$ and the auxiliary electrode at 2.6 V$_{rms}$, the focal length reaches a minimum of 3.3 mm while maintaining an 800 μm aperture. The auxiliary electrode optimizes the electric field distribution, enhancing the focusing effect. This LC lens array offers a low-cost solution with excellent tolerance characteristics for portable 2D/3D displays, with potential applications in AR and VR technologies.

**Funding**

This work is supported by the National Natural Science Foundation of China (62305033).

## Disclosures

The authors declare that there are no conflicts of interest related to this article.

## Data availability

Data underlying the results presented in this paper are not publicly available at this time but may be obtained from the authors upon reasonable request.